\documentclass[letterpaper,10pt]{article} 
\usepackage{opticameet3} 
\usepackage{amsmath}
\usepackage{tabularx}
\usepackage{wrapfig}
\newcommand\authormark[1]{\textsuperscript{#1}}
\usepackage{caption}
\usepackage{etoolbox}
\makeatletter
\patchcmd{\thebibliography}{\small}{\footnotesize}{}{}
\patchcmd{\thebibliography}{\normalsize}{\footnotesize}{}{}
\makeatother
\makeatletter
\renewcommand{\refname}{\small References} 
\makeatother

\usepackage{amsmath,amssymb}
\usepackage[colorlinks=true,bookmarks=false,citecolor=blue,urlcolor=blue]{hyperref} 
\usepackage{caption}
\usepackage{subfig}
\begin{document}

\title{Radio-Optical Confluence in Intelligent Edge Networks}


\author{\author{Akshita Gupta,\authormark{1} Devika Dass,\authormark{1} Agastya Raj,\authormark{1} Carlos Natalino,\authormark{2} Marco Ruffini,\authormark{1}
 Paolo Monti,\authormark{2} and Daniel Kilper\authormark{3}}
\address{\authormark{1} School of Computer Science and Statistics, Trinity College Dublin, Dublin, Ireland\\
\authormark{2} Department of Electrical Engineering, Chalmers University of Technology, Gothenburg, Sweden\\
\authormark{3}CONNECT, Trinity College Dublin, Dublin, Ireland\\}
\email{\authormark{} agupta1@tcd.ie, dassd@tcd.ie, rajag@tcd.ie, carlos.natalino@chalmers.se, mpaolo@chalmers.se, marco.ruffini@tcd.ie, dan.kilper@tcd.ie}}



\begin{abstract}
Challenges associated with densification of radio access networks are motivating exploration of more efficient and scalable architectures. We examine recent progress in one direction that involves moving beyond radio and optical convergence to full confluence.   
\end{abstract}

\section{Introduction}
\begin{wrapfigure}{r}{0.55\textwidth}
    \centering
    \includegraphics[width=\linewidth]{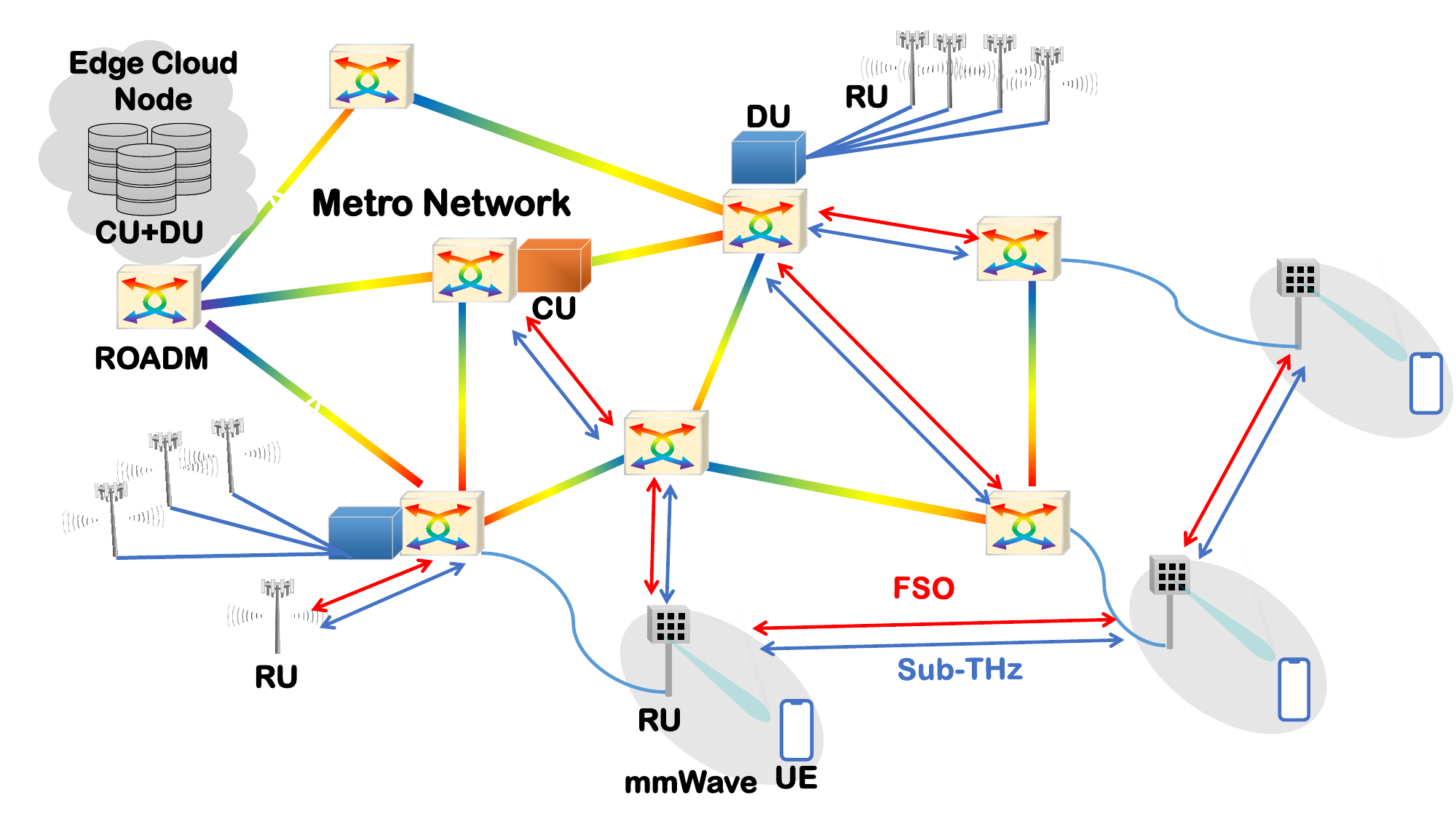}
    {\footnotesize
    \caption{Radio-optical confluent network model with fiber and wireless links between radio access points (RU/DU) and metro network nodes (ROADM).}
    \label{fig:vision_network}}
\end{wrapfigure}

Advancement in computing power, network capacity, and data storage have enabled bandwidth-intensive applications such as artificial intelligence, big data analytics and cloud computing. These applications have further contributed to both the continued growth of global internet traffic originating from the access network and increasing traffic between data centers, which most recently includes edge computing facilities \cite{itu_global_connectivity_report_2025}.
With the widespread rollout of 5G networks and the emergence of 6G specifications aimed at supporting even more demanding mobile use cases, optical networks — serving as the backbone infrastructure from the metro edge to the core — provide a potential scope for accessing enhanced performance to meet these demands. While optical technologies provide high capacity and low latency, they can be costly and inefficient for use in much of the network edge where traffic is bursty and access points can be difficult to reach with fiber.  One approach that was recently proposed is to use tighter convergence of optical and radio technologies to enable new modalities of dynamic and efficient operation. 

Confluent transmission~\cite{rishu} considers seamless interoperability and coordinated operation across diverse transport technologies combining, for example, radio fixed wireless (RFW) — including mmWave and sub-terahertz (sub-THz)/ THz bands—free-space optics (FSO), and flex-grid wavelength-division multiplexed (Flex-WDM) fiber-optic links within the same end-to-end infrastructure as shown in Fig.~\ref{fig:vision_network}. The use of wireless links enables the network to reach a greater density of access points and motivates the extension of mesh networks in the metro core all the way to the edge access points. This can be done over either passive optical network (PON) infrastructure or through adaptive reconfigurable optical add-drop multiplexers (ROADMs) or other new wavelength switching technologies~\cite{ofc2025_postdeadline}. The challenge in densification is the cost and energy sensitivity of these systems. The confluent approach is then attractive for two reasons: \emph{1)} to integrate the wired and wireless infrastructure avoiding duplicate systems and system handovers and \emph{2)} to allow for common high speed control and adaptation across the overall mesh network to support greater efficiency in response to changing traffic patterns. Understanding the potential impact of the reduced duplication aspects involves considering the densification costs from a mobile network deployment aspect. The benefits can also be evaluated in terms of the fiber network and its ability to deliver capacity and low latency to the edge. Using mesh networks to reduce the number of hops in the network provides lower latency, increased capacity (reduced blocking), and increased resilience. In addition, this infrastructure needs to transport combinations of wavelength multiplexed front-, mid-, and back-haul signals across each of the fiber and wireless links. In particular, the use of analog front-haul signals enables greater spectrum efficiency and simplification of the radio units. Thus, it is important to understand the transmission properties of analog radio over fiber (ARoF) signals multiplexed with digital coherent optical (DCO) signals. Ultimately, these transmission properties need to be taken into account in the transmission design of a confluent mesh network. 

In this paper, we present an analysis of recent progress on understanding the performance and costs associated with the use of radio–optical confluence to form mesh edge and metro networks. In particular, we describe recent studies that have considered the use of FSO and THz connections relative to fiber connections to support future 6G networks. These results indicate key cost and performance trade-off thresholds and show potential cost benefits from combinations of wireless and fiber to support high density radio access.  In addition, we present recent work on ARoF and DCO signals within metro-edge and access networks. Emerging multiplexing architectures enable these heterogeneous signals to efficiently share the same fiber infrastructure, thereby maximizing spectral utilization while maintaining signal integrity. We also look at the benefits of mesh connectivity at the edge when accounting for the analog and digital signal wavelength multiplexing constraints. Results indicate that with a small number of edge mesh links the average hop counts to a processing location can be significantly reduced, yielding mesh network benefits. Such developments further strengthen the feasibility of integrated radio–optical transport solutions, supporting scalable, flexible, and cost-efficient deployment of future 6G networks.

\section{Fiber and Wireless Mesh Networks}
In considering this new confluent network paradigm, we adopt the edge network design in Fig.~\ref{fig:vision_network}. The radio access network (RAN) portion is centralized with processing at one or more metro data center/edge cloud nodes. Both analog and digital radio over fiber signals are transported from the radio access point (RU) to either the centralized radio processing units (CU) or distributed radio processing units (DU). The same network that carries digital backhaul and digitalized fronthaul signals also supports the fronthaul transport of analog RoF signals. The edge cloud node performs both radio and application data processing. Note that FSO and sub-THz links are formed between RUs and/or metro ROADM nodes.  

Combining parallel FSO  and radio links, particularly in the case of mmWave or sub-THz/THz signals, has been studied in some detail and shown to provide resilience benefits due to the complementary tolerance to certain weather conditions. In recent work, combinations of FSO and sub-THz links were studied in the context of 6G access networks \cite{fso-thz}, where the impact of the network equipment deployment on network reliability and deployment costs was evaluated. This prior work explored dynamic resource allocation and on-demand network connectivity for point-to-point (P2P) and point-to-multipoint (P2MP) access networks. The key findings provide design guidelines for equipment vendors and network operators to deploy cost-efficient and resilient 6G access networks based on hybrid FSO–THz links, with or as an alternative to the fiber links.
In further work \cite{fiber-fso}, a hybrid fiber/FSO backhaul strategy for two-hop IAB networks was proposed, targeting 6G infrastructure planning. The study introduces a topology optimization algorithm that balances deployment cost against the spatial separation of fiber-connected nodes to maximize service coverage. Results confirm that hybrid architectures offer significant cost savings over fully fibered networks without compromising end-user coverage probability. Furthermore, the work highlights that intelligent, spatially-aware node selection is critical for maximizing energy efficiency while balancing the trade-off between throughput and power consumption.

\section{ARoF and DCO Transmission in Mesh Networks}

The multiplexing of analog and digital signals in confluent mesh networks can make use of a variety of strategies. In Fig. \ref{fig:wp4_OSAAS}, we show one such approach which allows for efficient banding of the ARoF signals with the DCO signals. Here, the ARoF signal distribution is modeled with 4 double-sideband ARoF signals of 500 MHz bandwidth and data rate of up to 3 Gbps. A total of about 12.5 GHz spectral window in the ROADM is allocated to one group of 4 RUs that can transmit the ARoF signals at 1, 2, 3, and 4 GHz. An ITU channel grid of 50 GHz could cater to a total of 16 RUs connected to one ROADM in an edge metro-access scenario. Our previous work has shown the feasibility of the coexistence of these analog and digital signals over multispan live-production metro-access networks, enabling heterogeneous transmission of up to 600 Gbps DCO traffic alongside a 6 Gbps 16-QAM orthogonal frequency division multiple access (OFDM) ARoF signal over a 77 km loop \cite{devika_jocn}. The DCO and ARoF channels, when centrally located over separate channels on the ITU grid, do not show measurable cross-interference. 
In particular, with Flex-grid WDM systems, we can integrate ARoF and digital signals with fine granularity to exploit the full optical spectrum. Previous experiments showed the use of multiple ARoF signals, constituting a narrowband signal resembling the 5G New Radio (NR) standard for mobile traffic and a wideband signal for broadband wireless fidelity (Wi-Fi) service, over a wavelength-flexible fronthaul~\cite{devika_multi_band}. Based on these results, the multiplexing scheme shown in  Fig. \ref{fig:wp4_OSAAS} is expected to support both the analog and digital signals across a metro network without appreciable degradation.

\begin{minipage}[t]{0.58\textwidth}
    \centering
    \includegraphics[width=0.9\linewidth]{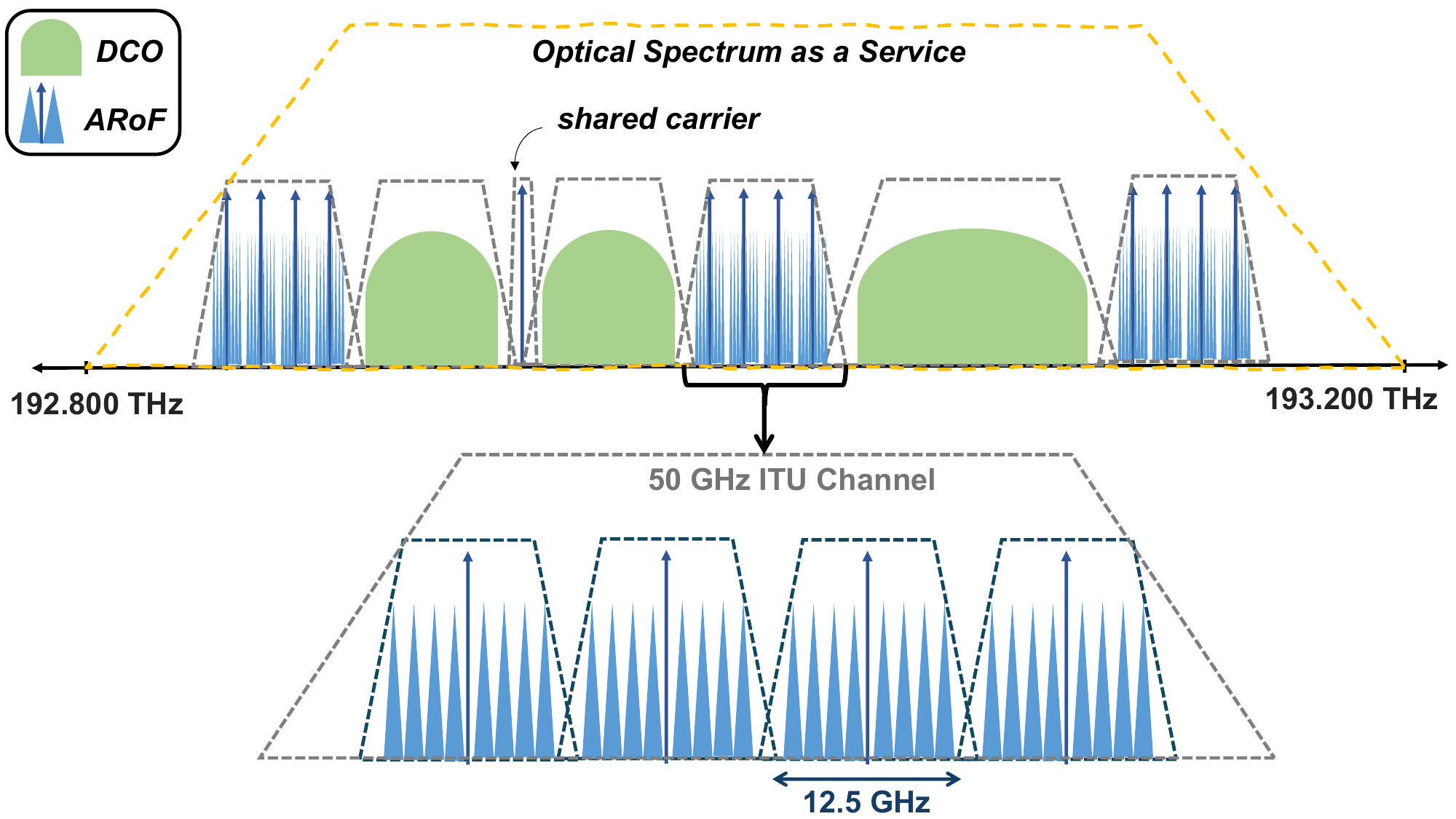}
\captionof{figure}{Example of an optical spectrum carrying wavelength multiplexed analog front-haul and digital baseband signals.}
\label{fig:wp4_OSAAS}
\end{minipage}\hfill
\begin{minipage}[t]{0.37\textwidth}
\centering
\includegraphics[width=1\linewidth,trim={0 6cm 0 5cm},clip]{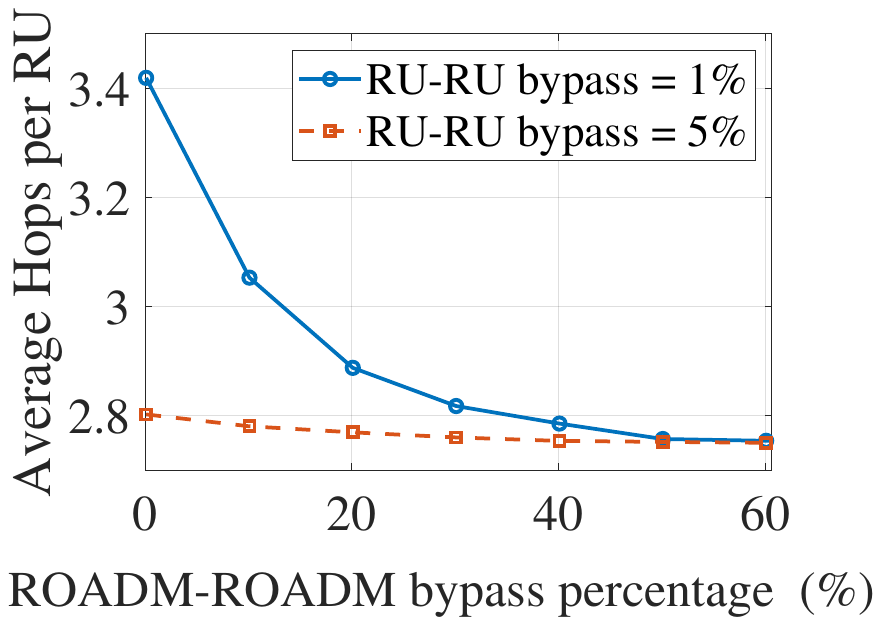}
    \captionof{figure}{Average hop count to the edge cloud with increasing mesh bypass connections between ROADMs or between RUs.}
    \label{fig:combined}
\end{minipage}

Building upon these physical-layer and architectural advancements, we explore mesh interconnectivity performance for an ARoF-based C-RAN architecture over the ROADM network in Fig. \ref{fig:vision_network}. In particular, we can explore how the constraints related to the multiplexing scheme in Fig. \ref{fig:wp4_OSAAS} influence the impact of mesh connectivity. As an example, we perform an analysis with a ring network of 10 ROADMs feeding into an edge cloud data center, and correspondingly, 16 RUs aggregated into each ROADM. Based on this model, we maximise the RU connectivity and analyse the impact of additional mesh-connection bypass links to minimise the number of hops in the network from any RU to the data center. We consider a centralized load-aware shortest-path routing algorithm designed for the confluent radio-optical mesh architecture. For each active RU, Dijkstra’s algorithm is used to compute the shortest path toward the nearest DC over the unified graph topology. After all candidate paths are identified, the network congestion is determined based on the load and the spectrum constraints over the different nodes in the network path. The benefits of introducing additional ROADM–ROADM bypass links are illustrated in Fig. \ref{fig:combined}, where with the increase in ROADM-ROADM connectivity, the average hops per RU decreases. Specifically, the average number of hops decreases from 3.4 to 2.8 hops per RU as the ROADM-ROADM bypass links increases from 0\% to 40\%. Further, with the increase in RU-RU bypass links, the average hops in the network decreases. Specifically, as the RU-RU bypass links increases from 1\% to 5\%, the average hops in the network decreases to approximately the same level as a fully connected mesh in the ROADM network. These results indicate that with just a small percentage of mesh cross-links to the RU access points, using for example FSO, one can achieve a large reduction in the average number of hops, approaching the performance of a highly connected mesh core network. Further studies are underway to fully explore the different trade-offs and constraints involved in confluent mesh networks and their transmission properties.

\vspace{-0.3cm}
\section{Conclusions}
This paper provides an evaluation of the performance and cost benefits of radio–optical confluence to form edge mesh networks in support of future mobile networks. Progress in understanding the benefits of wireless and fiber connections at the edge is reviewed, along with recent experiments on the transmission of ARoF and DCO signals in metro-edge networks. Further experiments are needed on transmission across fiber and wireless transport links, and their corresponding constraints will need to be incorporated into network design studies. Future work will also consider dynamic operation across the entire mesh network to determine whether latency and energy efficiency benefits can be realized.



\vspace{-7pt}
\section*{\small Acknowledgements} 
\vspace{-4pt}
\footnotesize{This work was supported by the European Commission projects ECO-eNET, which has received funding from the Smart Networks and Services Joint Undertaking (SNS JU) under grant no. 10113933, the OpenIreland Research Ireland grant 18/RI/5721, and the Research Ireland CONNECT Centre grant 13/RC/2077\_P2.}
\vspace{-6pt}

\bibliographystyle{IEEEtran}
\bibliography{main}

\end{document}